\documentclass{moriond}

\def\Journal#1#2#3#4{{#1} {\bf #2}, #3 (#4)}

\def\AA{{\em Astronomy and Astrophysics.} }

\def\be{\begin{equation}}
\def\ee{\end{equation}}
\def\bea{\begin{eqnarray}}
\def\eea{\end{eqnarray}}

\newcommand{\Photo}{}

\begin{document}
\vspace*{4cm}
\title{Neutron Stars Mass-Radius relationship and \\Electromagnetic follow-up of Kilonovae}

\author{  D. Barba Gonz\'alez, M. A. P\'erez-Garc\'ia}

\address{ Department of Fundamental Physics \& IUFFyM, University of Salamanca, E-37008 Spain}
\maketitle\abstracts{ When two Neutron Stars collide a multi-band electromagnetic emission, known as Kilonova (KN), follows being powered by the radioactive decay of ejecta products. In this contribution we discuss how the equation of state of dense matter, impacts the mass and velocity in the KN ejecta and thus its light curve. Using this information encoded in the stellar mass-radius relationship, we ellaborate on how the future experimental observations in photon channels, in addition to complementary multimessenger probes, could provide a new and more detailed insight into the equation of state of nuclear matter.}

\section{Introduction}

When two Neutron Stars (NSs) in a binary system, a BNS, coalesce and merge, they are expected to produce a multimessenger signal: gravitational waves (GW), photons from an electromagnetic (EM) counterpart and even neutrinos. One example of such event was the  observation of a rapidly fading EM transient in the galaxy NGC 4993, which was spatially coincident\cite{abbot} with GW170817. The optical and infrared emission, known as Kilonova (KN) or macronova, was named AT2017gfo. It was possible to identify line features in the spectra, consistent with the presence of light r-process elements (atomic masses of $A=90–140$) powering the emission. In addition, a gamma-ray burst GRB170817A, was also detected. Since then it is now expected that many detections of this kind or similar ones i.e. Black Hole and NS (BHNS) mergers will follow in the future. An updated BNS rate\cite{abbot}  after the discovery of GW170817  leads to the expectation of potentially detectable  dozens of events per year by LIGO/Virgo O4 and O5 runs. 

In order to capture the challenging  optical counterpart of a BNS merger, it is crucial to develop an alert system allowing  observers to point in the right direction in the sky, obviously before it has faded away and ideally before the time it peaks. The community has gained valuable experience thanks to the historical event GW170817 localised optically in a sky area of around 30 square degrees within hours of its appearance.  Typically, light curve peak values are obtained at times $t_{\rm peak}$ ranging between tenths of a day and a day. The alert system GraceDB\cite{grace}, a communication node that connects LIGO and Virgo analysis pipeline, sends an alert to astronomers when a promising binary GW signal has been detected and located in the sky. Although challenging, detecting these EM counterparts is key to shed light on some relevant physical processes taking place, such as dynamical and post-merger ejection, neutron capture and thermalization in the ambient matter. Existing and future missions, like for example MAAT \cite{maat} in the GTC (Spain) will be able to provide an agile response and deeper knowledge in this field. 
In this contribution we discuss how KNe can serve yet as another additional complementary constraint to the equation of state of neutron star matter, particularly through the light curve and spectra as measured by existing and future missions. 

\section{NS equation of state and key KN properties}

The  equation of state (EoS) of nuclear matter describes the behavior of the matter inside NSs. Typically these objects have a mass ($M$) up to a maximum value $M_{max}\sim 2 M_\odot$ and a radius ($R$) in the $10-13$ km range. Constraints on both quantities have been provided in a series of world-wide experimental efforts \cite{GWeos}. These objects appear as solutions of the stellar structure equations from General Relativity (GR) and their computed $M(R)$ values display some spread due to the poorly known high density matter EoS. In particular, it is not yet known what are the actual degrees of freedom suitable for its description, i.e. whether matter inside remains under the form of nucleons or perhaps deconfined as quarks. NSs are born in the aftermath of a Supernova explosion displaying a large velocity kick and distribute mostly towards the center of the galaxies. Further into the NS description, its layered structure can be mainly explained from two regions: an external crust and a central core. It is believed that up to nearly $99\%$ of the NS mass resides in the latter under the form of  nucleons, hyperons and alike species although if high  densities are in excess of $\sim 6 \times 10^{14}$ $\rm g/cm^3$ a quark deconfined state may appear\cite{olek1}.  From the population studies of Galactic double NSs (DNSs) it appears that $\simeq98\%$ of all merging DNSs will have a mass ratio, $q$, close to unity\cite{zhang} $q<1.1$. Recent studies\cite{dns} find that the weighted mean masses of the primary and companion stars are respectively $(1.439\pm0.036) M\odot$ and $(1.239\pm0.020) M\odot$. In addition, the surface magnetic field strength in the primary stars of the DNSs is $B\sim 10^{10}$ G, and the spin period is $P\sim 50$ ms. Due to the fact that the NS in the BNS events under scrutiny will share these general characteristics it is important to see what information the EM counterparts can  yield. 
The gravitational and baryonic masses are two quantities that determine the binding energy (BE) of the NS. Namely, the gravitational mass is  $m_G(R)=M=\int_{0}^{R} 4 \pi r^{2} \epsilon(r) d r$ where $\epsilon(r)$ is the energy density at a radial distance $r$ from the center and the  baryonic mass $M_B=M^{*}=m_{0} \int_{0}^{R} 4 \pi r^{2}\left[1-2 G m_{G}(r) / r c^{2}\right]^{-1 / 2} n(r) d r$ where
$m_{0}$ is the mass of a baryon and $n(r)$ is the baryon (nucleon) number density. The binding energy of the NS is $B E=\left(M^{*}-M\right) c^{2}$ and  typically $BE=|BE|$ is taken as a positive quantity. 

In the literature, some universal relations between the BE and $C$ have been proposed, such as that\cite{chinos} based on the radius of a $1.4 M_\odot$ NS, $R_{1.4}$, $M^{*}=M+R_{1.4}^{-1} \times M^{2}$, with $1.8 \%$ relative error. Note that the $1.4 M_\odot$ selection is close to the most probable for primary companions in DNSs.  In order to illustrate our findings we consider two EoS i.e. DD2  and SFHo \cite{sfho}, whose M(R) relationship is shown in Fig.(1)  left panel. For SFHo it is found a saturation density $n_0=0.1583$ $\rm fm^{-3}$, nuclear incompressibility $K=245.4$ MeV, symmetry energy $J=31.57$ MeV and a logarithmic derivative of the symmetry energy $L=47.10$ MeV. For the DD2  $n_{0}=0.1491\mathrm{fm}^{-3}$, $K=242.7$ MeV, $J=31.78$ MeV and $L=55.19$ MeV. Using existing data\cite{dietrich} from BNS simulations  we obtain $R^{\rm DD2}_{1.4}=13.5 \pm 0.1$ km, $R^{\rm SFHo}_{1.4}= 11.8 \pm 0.1$ km, see Fig.(1)  right panel. Considering the sum of BE of both NSs described by a given EoS, $BE_1+BE_2$, we arrive at values for DD2: $BE_1+BE_2=0.25\pm0.04M_{\odot}$ and  SFHo:  $BE_1+BE_2=0.28\pm 0.04M_{\odot}$ which fulfill the fitted form\cite{shao} $\frac{BE_i}{M_i}=-0.0130+0.618 C_i+0.267 {C^2_i}$ where $C_i=\frac{G M_i}{R_i c^2}$ is the compactness. They retain the $M(R)$ relationship from $C$ and $M$ values.

As shown in GR simulations\cite{bauswein} the ejected mass in a violent BNS merger is mostly stripped from the inner and outer crust of the NSs and will be determined by the energetics of the collision and the actual amount of matter in the crust. In a NS the neutron rich crust displays a large set of inhomogenous substructures ranging from lattice nuclei to {\it pasta phases} \cite{pasta}. In the thin-crust approximation $M_{\rm {crust }} \approx \frac{4 \pi R^{3} P_{t}}{c^2 C}[1-2 C]$ where $P_t$ is the EoS dependent pressure transition from the crust-core regions \cite{chamel}. There is a large sensitivity to the EoS in the crust mass value, being smaller for heavier stars. Recent Bayesian analysis\cite{carreau} provide values for the  transition density  $n_t = 0.072\pm 0.011$ $\rm fm^{-3}$ and the transition pressure $P_t = 0.339\pm 0.115$ $\rm MeV$ $\rm fm^{-3}$. The resulting crust masses $M^{\rm DD2}_{\rm {crust }}=0.038M_\odot$ and $M^{\rm SFHo}_{\rm {crust }}=0.024M_\odot$.  In addition, the number of baryons must be conserved $N^*_1+N^*_2=N^*_{remnant}+N^*_{ejec}$.

\begin{figure}[t]
\begin{minipage}{0.43\linewidth}
\centerline{\includegraphics[width=1.1\linewidth]{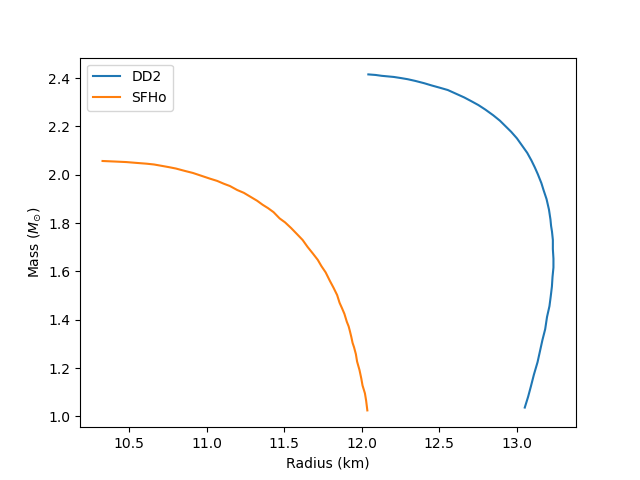}}
\end{minipage}
\hfill
\begin{minipage}{0.43\linewidth}
\centerline{\includegraphics[width=1.1\linewidth]{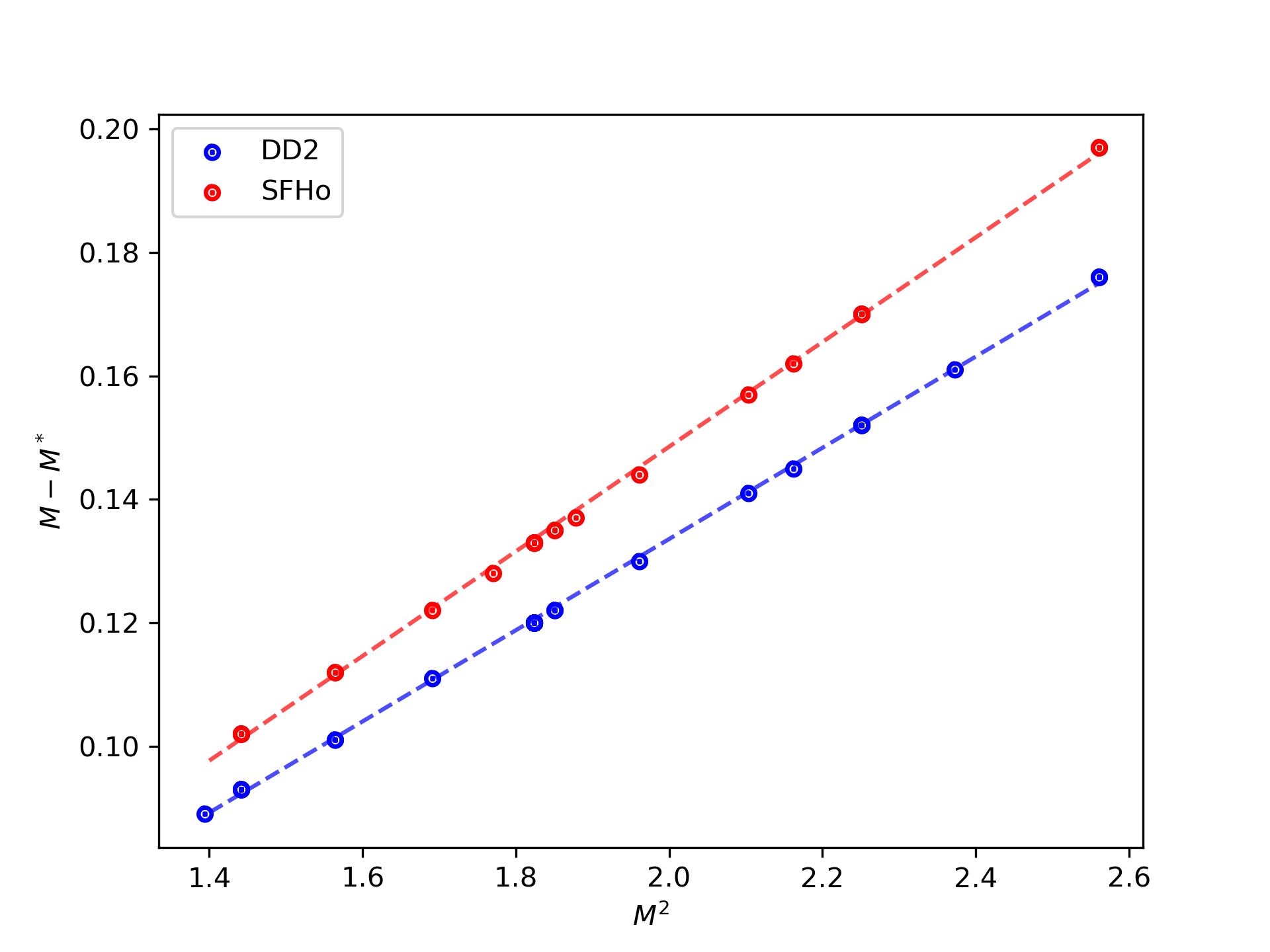}}
\end{minipage}
\caption[]{M(R) relationship for the  DD2  and SFHo EoS (left) and BE as a function of $M^2$ (right)}
\label{fig:panel}
\end{figure}

\begin{figure}[t]
\begin{minipage}{0.45\linewidth}
\centerline{\includegraphics[width=1.1\linewidth]{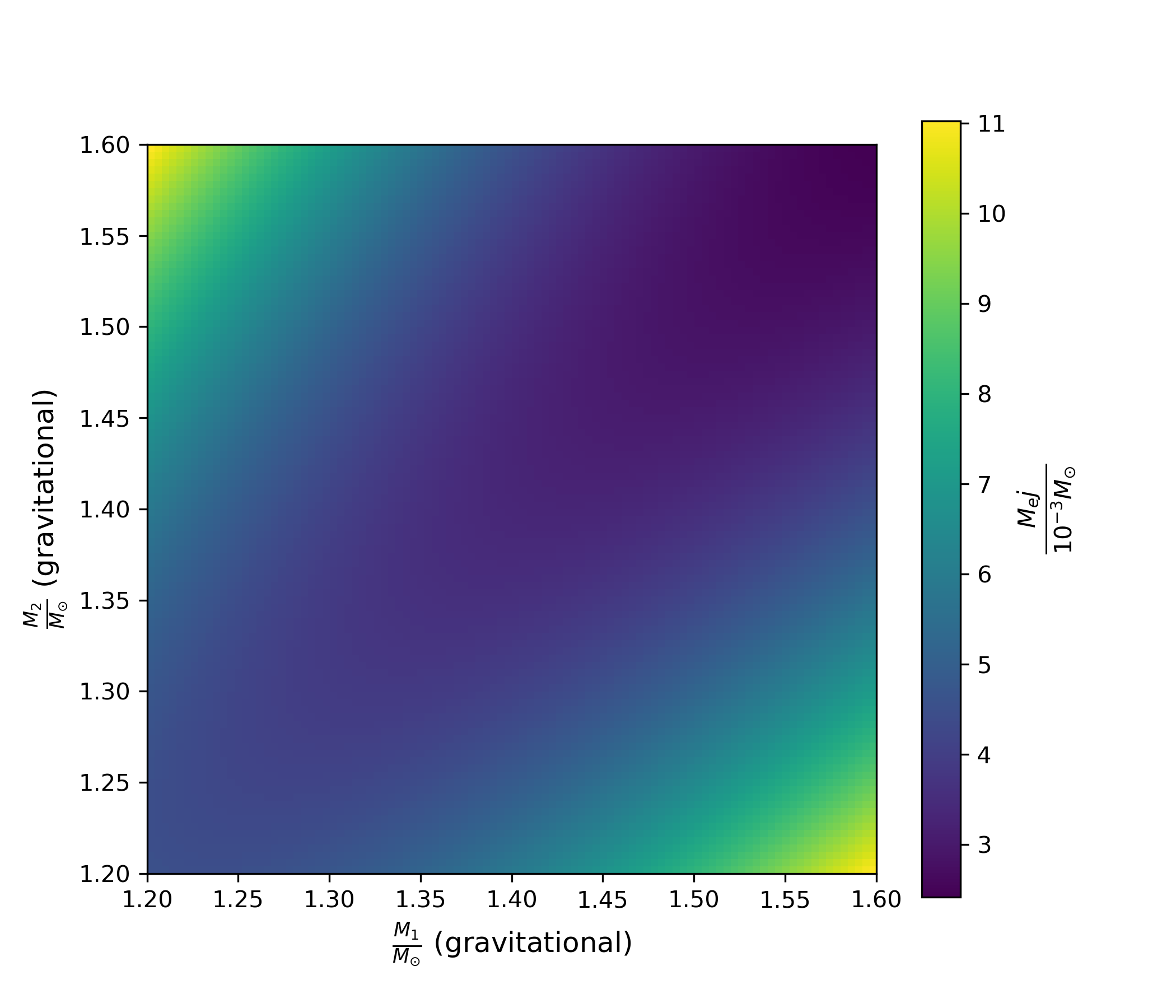}}
\end{minipage}
\hfill
\begin{minipage}{0.53\linewidth}
\centerline{\includegraphics[scale=0.27]{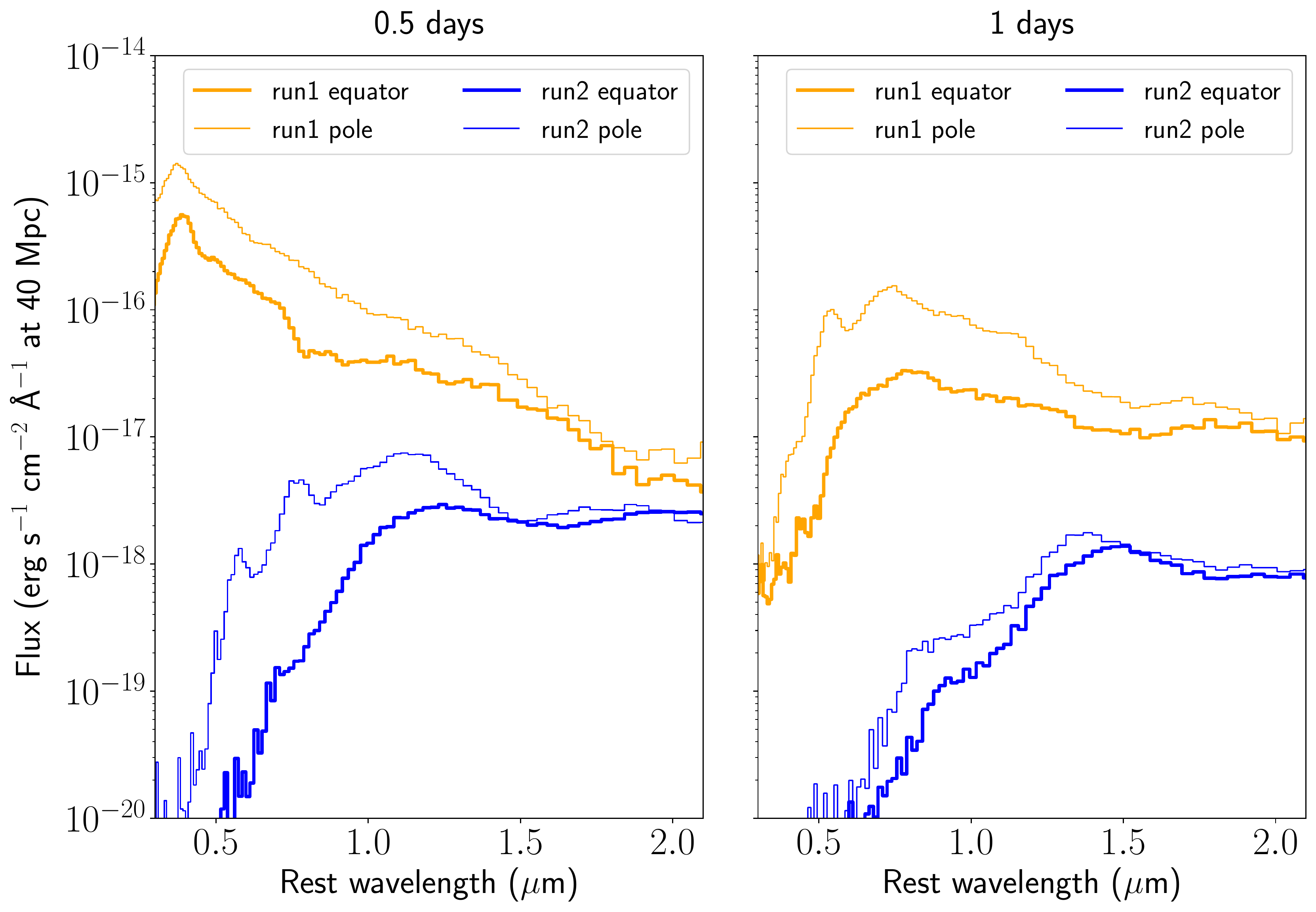}}
\end{minipage}
\caption[]{(left) Ejected mass from simulation data \cite{dietrich}. (right) Simulated flux for a BNS merger using DD2 (run 1) and SFHo (run 2) EoS at 40 Mpc for polar and equatorial view angles at 0.5 d (left) and 1 d (right) after the merge.}
\label{fig:spectra}
\end{figure}

The complex dynamics of the BNS merger and its ejecta are studied using GR numerical simulations with a number of approximations regarding neutrino transport and microphysics. The output is obtained by determining the relevant parameters modelling the KN emission, including the mass, $M_{\mathrm{ej}}$, and velocity, $v_{\mathrm{ej}}$, of the ejecta, which are fitted\cite{dietrich} (with some uncertainty, see parameterizations) to functional forms depending on both objects in the BNS ($1,2$),
\begin{equation}
\scriptstyle
\frac{M_{\mathrm{ej}}^{\mathrm{fit}}}{10^{-3} M_{\odot}}=\left[a\left(\frac{M_{2}}{M_{1}}\right)^{1 / 3}\left(\frac{1-2 C_{1}}{C_{1}}\right)+b\left(\frac{M_{2}}{M_{1}}\right)^{n}+c\left(1-\frac{M_{1}}{M_{1}^{*}}\right)\right] M_{1}^{*}+(1 \leftrightarrow 2)+d,
\end{equation}
\begin{equation}
v_{\mathrm{ej}}=\sqrt{v_{\rho}^{2}+v_{z}^{2}} , \quad v_{\rho, z}=\left[a_{\rho, z}\left(\frac{M_{1}}{M_{2}}\right)\left(1+c_{\rho, z} C_{1}\right)\right]+(1 \leftrightarrow 2)+b_{\rho, z}.
\end{equation}
We plot in Fig.(\ref{fig:spectra}) left panel, the ejected mass as a function of gravitational masses $M_1, M_2$ for $M_i \in [1.2,1.6]$ in our analysis with the DD2 EoS. As a rule of thumb we see that smaller mass ratio $q\equiv M_</M_>$ provide larger $M_{ej}$. The experimental measurement of the KN peak bolometric luminosity and spectral time series usually depend on the conditions of transparency after the ejection of matter parameterized  through the opacity, $\kappa$. For early emission, typically $\kappa \sim {0.1 \mathrm{~cm}^{2} \mathrm{~g}^{-1}}$  while for later r-powered emission $\kappa$ could be a factor of 10  or more larger.  The peak values and are given \cite{tanaka} by $t_{{peak}}=4.9 \mathrm{~d} \times \left(\frac{M_{ej}}{10^{-2} M_{\odot}}\right)^{\frac{1}{2}}\left(\frac{\kappa}{10 \mathrm{~cm}^{2} \mathrm{~g}^{-1}}\right)^{\frac{1}{2}}\left(\frac{v_{\mathrm{ej}}}{0.1}\right)^{-\frac{1}{2}}$ .
 For a BNS the measurable magnitudes can be fitted to the Black Body power spectrum \cite{abbot} and from GW170817 some constraints appear for masses i.e. total mass $M \approx 2.74 M_{\odot}$, and individual masses of $M_{1} \approx(1.36-1.6) M_{\odot}$ and $M_{2} \approx(1.17-1.36) M_{\odot}$. Also from this analysis marginalized over the selection methods, it was obtained that the radius of the progenitors $R_{1} \sim(10.8-11.9) \pm 2 \mathrm{~km}$ and $R_{2} \sim(10.7-11.9) \pm 2 \mathrm{~km}$ and the tidal deformability parameters $\Lambda_{1}<\Lambda_{2}, \Lambda_{1}<500$ and $\Lambda_{1}<1000$. Under this additional input they find $\tilde{\Lambda}(310-345)_{-245}^{+691}$ and $R_{1.4}=8.7-14.1 \mathrm{~km},$ where the binary tidal deformability $\tilde{\Lambda}$ is defined as $
\tilde{\Lambda} \equiv \frac{16}{13}\left[\frac{\Lambda_{1} M_{1}^{4}\left(M_{1}+12 M_{2}\right)}{\left(M_{1}+M_{2}\right)^{5}}+1 \leftrightarrow 2\right]$ which is most precisely derived.

Direct measurements from current and future missions will be able to provide time spectral series. In Fig.(\ref{fig:spectra}), right panel, we plot  the simulated fluxes of two runs using POSSIS \cite{possis} as a function of the EM rest wavelength. Run 1 (Run 2) corresponds to a $q=1$ BNS merger of $1.4M_\odot$ at 40 Mpc using the output of recent simulations\cite{radice} $M140140-LK$ of DD2 (SHFo) EoS using polar/equatorial orientations. We notice that the early, agile KN follow up would be key to further constrain the EoS. The soft SFHo EoS peaks at IR, and is fainter than DD2 across the whole spectrum. Since the $q$ values remain the same we rely to the EoS features (including secondary peak features) for wavelengths less than $\sim 1\mu$m. 

To summarize, we have discussed the dependencies of the EM observables obtained from KN light curves (peak values) with $q=1$ in order to help indirectly constrain quantities such as the $M(R), C, BE, R_{1.4}$ and thus the EoS of NS matter. In particular, the combination of optical/IR observations from proposed missions such as MAAT in addition to GW signals from the LIGO/Virgo collaboration, will provide complementary valuable input. Further work is in progress.

\section*{Acknowledgments}

We acknowledge  useful comments from C. Albertus, M. Bulla, L. Izzo and F. Prada.  Spectra in Fig. (2) are courtesy of M. Bulla. This research has been supported by University of Salamanca, Junta de Castilla y Le\'on  SA096P20, Spanish Ministry of Science  PID2019-107778GB-100 projects, MULTIDARK Spanish network  and PHAROS COST Action CA16214.

\end{document}